\documentclass[lettersize,journal]{IEEEtran}
\IEEEoverridecommandlockouts
\usepackage{algorithmic}
\usepackage{algorithm}
\usepackage{orcidlink}
\usepackage[shortlabels]{enumitem}
\usepackage{subcaption}
\usepackage{array}
\hypersetup{%
    pdfborder = {0 0 0}
}
\usepackage{cuted}

\usepackage{textcomp}
\usepackage{stfloats}
\usepackage{url}
\usepackage{amsmath,amsfonts}
\usepackage{verbatim}
\usepackage{graphicx}
\usepackage{cite}
  \makeatletter
\newcommand*{\rom}[1]{\expandafter\@slowromancap\romannumeral #1@}
\makeatother
\begin{document}
 \bstctlcite{IEEEexample:BSTcontrol}

\title{{\color{black}{Grover-Based PLS: AUD and Beamforming with Artificial Noise in CD-NOMA}}}

 \author{Deemah H. Tashman,~\IEEEmembership{Member, IEEE}, and Soumaya Cherkaoui,~\IEEEmembership{Senior Member, IEEE} 
\thanks{D. Tashman and S. Cherkaoui are with   LINCS Laboratory, Department of Computer  and Software Engineering, Polytechnique Montreal, Montreal, QC, Canada,  H3T 1J4 (e-mail: \{deemah.tashman, soumaya.cherkaoui\}@polymtl.ca). LINCS Lab website: https://lincslab.ca/en/} 
 }



\maketitle

\begin{abstract}
Sixth-Generation (6G) networks will require massive connectivity, ultra-low latency, and robust security, making reliable Active User Detection (AUD) essential for interference control and physical layer protection. This letter proposes a {\color{black}Grover-based} physical layer security (PLS) framework for a code-domain non-orthogonal multiple access (CD-NOMA) network, where the base station employs artificial-noise (AN)-assisted beamforming and identifies the active set via Grover's quantum search algorithm. We consider two threat models: passive eavesdroppers formed by detected inactive users, and active eavesdroppers selected as the top $f\%$ most frequent transmitters among detected active users. By aligning beams and AN with the Grover-based AUD output, the proposed scheme enlarges the main–wiretap rate gap and significantly improves the average secrecy rate compared with compressive sensing and classical correlation receiver baselines, while approaching maximum-likelihood detection performance with a quadratic reduction in search complexity. The impact of the information/AN power split, the base station transmit power, and the fraction of highly active users treated as eavesdroppers on secrecy is characterized through numerical simulations, and design insights are extracted for 6G PLS under both passive and active eavesdropping.
 \end{abstract}


\section{Introduction}

\IEEEPARstart{P}{hysical} layer security (PLS) is expected to play a central role in Sixth-Generation (6G) wireless networks, where massive connectivity and heterogeneous services exacerbate the risk of eavesdropping. PLS guaranties confidentiality by ensuring that the capacity of the legitimate (main) link exceeds that of the wiretap link, typically by enhancing the desired channel and/or degrading adversarial channels. Among available techniques, artificial-noise (AN) assisted beamforming has proven effective, as it steers information signals toward legitimate users while injecting AN in the null space of their channels, thereby jamming unauthorized receivers with limited interference to intended users~\cite{sharma2022physical}. The effectiveness of such PLS schemes critically depends on accurately identifying the active legitimate users and likely eavesdroppers. In grant-free code-domain non-orthogonal multiple access (CD-NOMA) system, user activity must be inferred from noisy superpositions of code-domain signatures, and the search over activity patterns grows combinatorially with the number of potential users, making conventional active user detection (AUD) methods increasingly complex and error-prone.

 Quantum search offers an attractive alternative by exploiting superposition and parallelism to reduce search complexity while maintaining high detection accuracy. For instance, Grover quantum search algorithm and traditional algorithms were contrasted in \cite{9833942} for the AUD process in a   NOMA system, focusing  exclusively   on the impact of noise. In \cite{9673141}, a comparable investigation was conducted with regard to optical code division multiple access, where the Durr and H{\o}yer quantum search technique was implemented to identify active users in wireless systems. {\color{black} The authors in \cite{10486901} proposed quantum minimum-searching algorithms for AUD in wireless IoT networks by adapting Durr and H{\o}yer-based search techniques and exploiting prior system knowledge to reduce complexity. Moreover, in \cite{11161301}, the authors proposed an enhanced BBHT-based Grover search   for AUD   by introducing dynamic distributions for the search parameters in order to reduce computational complexity. Recently, Grover-based AUD was employed to detect active secondary users in a cognitive radio network, taking into account energy constraints in \cite{11059714}.} {\color{black} In addition, the authors in \cite{piron2024quantum} and \cite{10279631} employed quantum annealing (QA) for AUD and multi-user detection (MUD), respectively, demonstrating that QA naturally maps wireless detection problems into quadratic unconstrained binary optimization (QUBO)/Ising optimization formulations. QA may be more compatible with currently available annealing hardware; however, unlike Grover-based search, QA does not generally provide the same provable quadratic query-complexity speedup guarantees under an ideal oracle.} 

{\color{black}Compared to the aforementioned works and to the best of our knowledge, this is the first work to pair a quantum search–based AUD mechanism  for enhancing PLS in a CD-NOMA network. Prior quantum-aided AUD studies focused on activity-detection accuracy, without considering secrecy metrics or AN-assisted beamforming. In contrast, we embed Grover-based AUD in a CD-NOMA PLS framework with AN-assisted beamforming, evaluate the resulting secrecy rates, and introduce a threat model that distinguishes passive eavesdroppers (detected inactive users) from active eavesdroppers (selected as the most frequently detected transmitters), thereby tying the eavesdropper behavior to the AUD dynamics.}
The main contributions are  given as follows:

\begin{itemize}
  \item  We develop a PLS scenario for a CD-NOMA network  that combines beamforming with AN to improve the main link and disrupt wiretap links. We adopt CD-NOMA as  it multiplexes multiple users on the same resources via distinct code signatures, enabling massive grant-free access that our beamforming/AN and AUD can exploit. {\color{black}{CD-NOMA is adopted over power-domain NOMA   and orthogonal multiple access techniques since its code-domain spreading naturally fits the proposed Grover-based activity-pattern search framework, avoids the successive interference cancellation-related limitations of power-domain schemes, and supports massive grant-free connectivity through user overloading.}}

\item We formulate and implement AUD using Grover’s method to determine the active legitimate set, considering untrusted users as potential eavesdroppers, which facilitates beam/AN alignment to the correct set. That is, we examine two types of eavesdroppers: passive eavesdroppers (detected inactive) and active eavesdroppers (top $f\%$ of most-frequent transmitters among detected actives).

\item We quantify the complexity--performance tradeoff of the proposed Grover-based AUD against multiple baselines, showing that Grover approaches maximum-likelihood (ML)  secrecy performance while achieving a substantially lower search complexity. 
\end{itemize}

\section{System Model}
 \begin{figure}[b]
  \centering
  \includegraphics[width=0.7\linewidth]{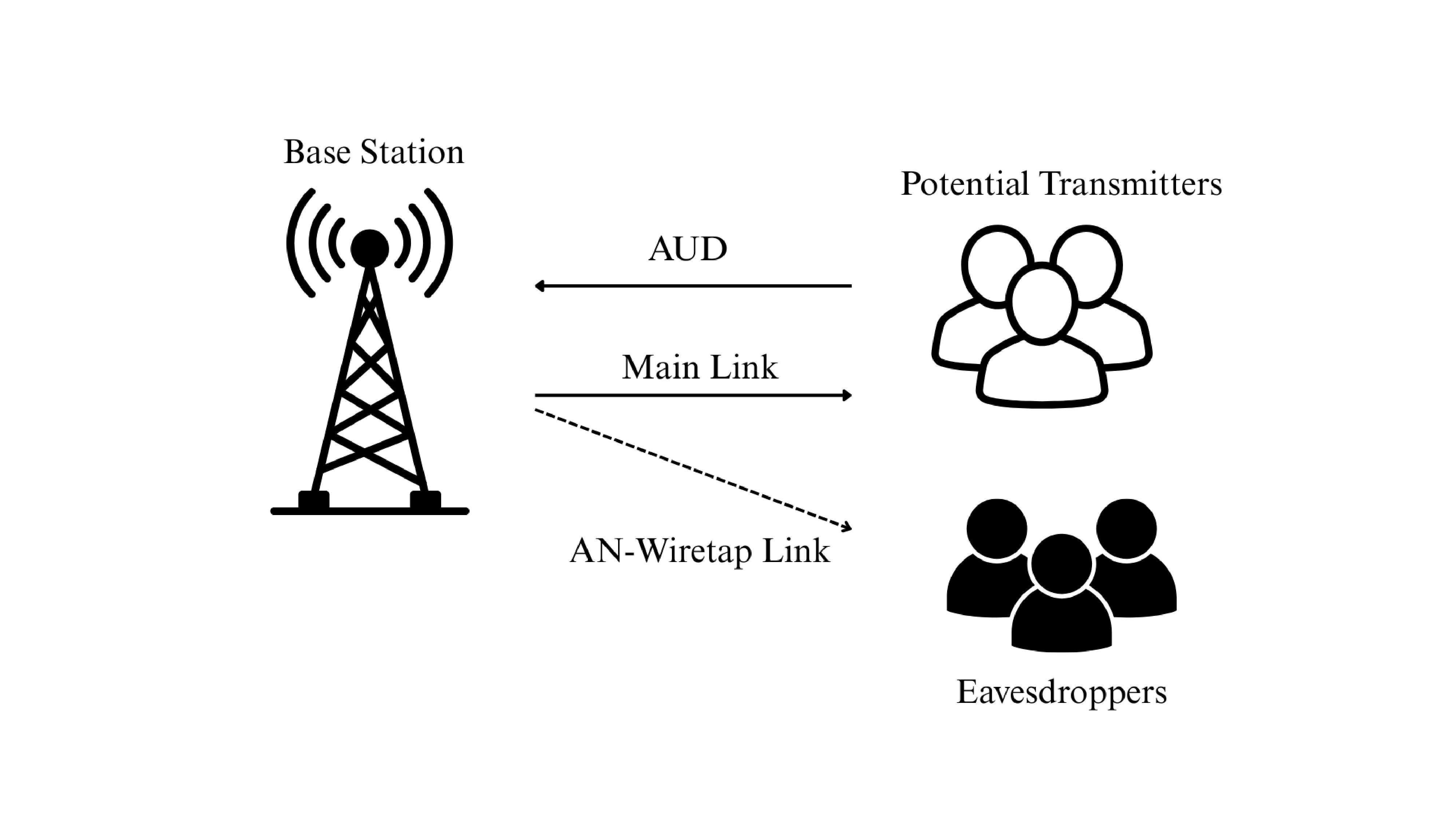}
  \caption{The system model.}
  \label{sys1}
\end{figure}
We presume that $N$ users are engaged in communicating with a base station (BS) equipped with $A$ antennas, as illustrated in Fig. \ref{sys1}.  Active users will transmit their uplink messages to the BS using the channel $\mathbf{h_{UL,k}}$ for $k \in \{1,2,\cdots,K\}$, where $K$ is the total number of active users.   The users utilize a CD-NOMA approach, wherein each employs a distinct code $C_n$, for $n \in \{1,2,\cdots,N\}$, comprising $\kappa$ chips (spreading factor), to spread the sent data.  The activity status of each user is denoted by $a_n$, such that
\begin{IEEEeqnarray}{lCr} 
a_n= \begin{cases}
   \text {1}    , & \text{user} \; n \; \text{is active}  \\
    \text {0} , &  \text{user} \; n \; \text{is idle}
 \end{cases} . 
\end{IEEEeqnarray} 
The received signals from all  users are accumulated at  BS as
\begin{IEEEeqnarray}{lcr}\label{y_bs}
y_{BS}= \sum_{n=1}^{N} \sqrt{P_n} \mathbf{h_{UL,n}} a_{n}  +n_{BS},
\end{IEEEeqnarray}  
\noindent where   $P_n$ is user $n$ transmission power for $n \in \{1,2,\cdots, N\}$, and $n_{BS}$  is the AWGN at  BS with 0 mean  and a variance of $N_0$.   The BS   uses the received messages to decide which users are active  and which are not using Grover's search algorithm. 

\noindent After deciding on the activity status of the $N$ users, the BS will use the beamforming technique to forward the useful messages to the legitimate users, while transmitting AN towards the eavesdroppers.   We consider two eavesdropper regimes aligned with practical threat cues depending on the outcome of Grover's search. \textit{First}, idle users are treated as passive eavesdroppers or untrusted devices. That is, despite not transmitting, they can still overhear downlink traffic; hence,  labeling them as eavesdroppers is a conservative, robustness-oriented assumption \cite{10278964}. This scenario can occur when the BS has no control over inactive users' behavior, and they could be in range to intercept the signals, posing a potential security threat. {\color{black}{\textit{Second},  among the detected active users, we classify the top \(f\%\) that exhibit persistent presence across detection intervals as active eavesdroppers. Such persistence is a well-recognized indicator of insider or stealthy adversaries attempting to remain embedded in the network state to bias channel estimation or monitor communications~\cite{8228655}.   Therefore, users exhibiting persistent transmission behavior can be considered anomalous and potentially malicious, since persistent transmission  allows an adversary to continuously update its estimate of the legitimate users' channels and data symbols. This increases  its interception capability over  time, while also reducing the null-space dimension available for AN transmission. Hence, selecting the top \(f\%\) most frequent transmitters models stronger insider adversaries.}} Evaluating both passive and active eavesdropping cases allows us to assess the robustness of the proposed PLS framework under both stealthy and persistent threats. Since the eavesdroppers remain network users, their CSI is naturally acquired through routine uplink pilot signaling at the BS, even if they are considered untrusted~\cite{10278964}. The received message  at the $k^{th}$ legitimate receiver  is given as
\begin{IEEEeqnarray}{lcr}\label{y_k}
  y_k = \mathbf{H_k} \mathbf{m_k} \sqrt{P_{\text{info}}} s_k + n_k,
\end{IEEEeqnarray}
\noindent Where $ \mathbf{H_{k}} \in \mathbb{R}^{A \times 1} $ is the channel vector between the BS and the $k^{th}$ legitimate user, $\mathbf{m_k}$ is the beamforming vector for the legitimate users, $s_k$ is the data symbol for the  $k^{th}$ legitimate user, \( P_{\text{info}} \) is the power allocated to the legitimate users, and $n_{k}$ is the AWGN at the $k^{th}$ legitimate user with 0 mean and a variance of $\sigma_k^2$ .  To prevent interference at legitimate users, the jamming beamforming vector $\mathbf{z_e} \in \mathbb{C}^{A \times 1}$ is designed to lie in the null space of the stacked channel matrix $\mathbf{H}_{\text{active}} \in \mathbb{C}^{K \times A}$ \cite{sharma2022physical}, where:
$\mathbf{z_e} \in \text{null}(\mathbf{H}_{\text{active}})$.
This ensures $\mathbf{H_k} \mathbf{z_e} = 0$ for all $ k \in \{1, \dots, K\}$; thus, the jamming signal does not affect legitimate receivers. Given (\ref{y_k}), the signal-to-noise ratio (SNR) at each legitimate user $k$ is given by
\begin{IEEEeqnarray}{lcr}
\gamma_k = \frac{P_{\text{info}}   | \mathbf{H}_k   \mathbf{m}_k |^2  }{   \sigma_k^2} .
\end{IEEEeqnarray}
\noindent  The   eavesdroppers are denoted by $E$ and the received signal at the $e^{th}$ eavesdropper is expressed as
\begin{IEEEeqnarray}{lcr} \label{y_e}
  y_{\text{e}} = \mathbf{H}_{\text{e}} \mathbf{m_k} \sqrt{P_{\text{info}}} s_k + \mathbf{H}_{\text{e}} \mathbf{z_e} \sqrt{P_{\text{jam}}} s_j + n_{\text{e}},
\end{IEEEeqnarray}
\noindent where  $\mathbf{H_{\text{e}}}\in \mathbb{R}^{A \times 1} $ is the channel vector between the BS and the $e^{th}$ eavesdropper,     \( P_{\text{jam}} \) is the power allocated to the jamming signal $(s_j)$, and $n_e$ is the AWGN   added to the received signal of 0 mean and a variance of $\sigma_e^2$ . We assume that the BS transmission power is denoted by $P_{BS}$ and is divided between $P_\text{info}$ and $P_{\text{jam}}$ according to a splitting factor $(\alpha)$, such that $P_{\text{info}}=\alpha P_{BS}$ and $P_{\text{jam}}=(1-\alpha) P_{BS}$.
For a worst-case scenario assumption, we consider that the eavesdroppers are colluding in the interception, where they  jointly process the gathered intercepted information by sending it to a centralized processor. Hence, we can replace the   multiple colluding eavesdroppers  by an eavesdropper of $E$ antennas that utilizes  maximum ratio combining (MRC) technique over the received signals \cite{9094381}. Hence, the signal-to-interference-plus-noise ratio (SINR) is calculated using (\ref{y_e}) as
\begin{IEEEeqnarray}{lcr}
\gamma_{e}^{\mathrm{coll}} =
\frac{\sum\limits_{e \in  {E}} P_{\text{info}} \left| \mathbf{H}_{e}\mathbf{m}_k \right|^2}
{\sum\limits_{e \in  {E}} P_{\text{jam}} \left| \mathbf{H}_{e}\mathbf{z}_e \right|^2 + | {E}|\,\sigma_e^2},
\end{IEEEeqnarray}


We assume that all channels follow the Rayleigh fading model. Accordingly, the corresponding channel power gains are exponentially distributed with fading parameter $\lambda_l$, where $l \in \{\mathbf{h}_{UL,n}, \mathbf{H}_k, \mathbf{H}_e\}$.

\section{{\color{black}{Grover-based AUD Algorithm \& PLS Analysis}}}
\subsection{Grover-Based AUD Framework: Oracle \& Diffuser Design}
During the first transmission phase, active users transmit pilot or data signals to the BS, whose objective is to accurately identify the active set in order to enable proper beamforming.    Utilizing quantum features, such as superposition, the BS may simultaneously evaluate multiple user activity states to reduce the computational complexity of the AUD process. For instance, Grover's algorithm is regarded as an effective method for identifying specific values in an unsorted database of size $D$. {\color{black}{That is, Grover’s algorithm employs qubits, which can exist in a superposition of the \(|0\rangle\) and \(|1\rangle\) states, in contrast to classical bits that represent either \(0\) or \(1\) \cite{nielsen2010quantum}.}}  This enables parallel computations and enhances the search algorithm's efficiency \cite{grover1996fast}. 
Moreover, conventional search algorithms have a complexity of $\mathcal{O}(D)$, while Grover's algorithm has a complexity of $\mathcal{O}(\sqrt{D})$, such that $D=2^N$. This leads to exploring multiple states concurrently due to the utilization of quantum superposition, speeding up the results. {\color{black}{We note that the reported \(\mathcal{O}(\sqrt{2^N})\) complexity refers to the ideal Grover query complexity over the activity-pattern search space.}}

Grover's algorithm is comprised of the Oracle and Diffuser components \cite{9833942}.  The Oracle's objective is to identify the state that corresponds to a specific value $\eta$, which denotes the actual received signal, i.e., the transmitting user.  In other words, the Oracle function $(O_\eta)$ modifies the phase of the state $|x\rangle$ to $-1$ when the function $Y(x)$ matches the predetermined value $(\eta)$, otherwise, the state remains unchanged.   {\color{black}{This is achieved by utilizing an auxiliary qubit initialized in the state
\[
|-\rangle = \frac{|0\rangle - |1\rangle}{\sqrt{2}},
\]
which enables the oracle to perform phase inversion on the marked state.}}   The Oracle component's procedure can be characterized as
\begin{IEEEeqnarray}{lCr} \label{oracle}
\textbf{Oracle: }O_\eta|x\rangle= \begin{cases}
   -|x\rangle   , & \text{if} \; Y(x)=\eta  \\
   |x\rangle  , &  \text{if} \; Y(x) \neq \eta 
 \end{cases}.
\end{IEEEeqnarray} 
 {\color{black}{Specifically, the Oracle process is performed using four registers. First, the index register is initialized in a superposition of all possible user activity patterns, where each basis state represents a candidate activity vector \(\mathbf{a}\in\{0,1\}^N\). Second, the value register accumulates the corresponding CD-NOMA codeword contributions associated with each candidate pattern. Third, the reference register stores the quantized version of the received signal, which already reflects the effects of the users' transmit powers, channel coefficients, and noise before quantization. Finally, the mark register inverts the phase of the candidate state when the accumulated value matches the quantized received observation, thereby marking the valid activity pattern for amplification by the Diffuser.}}  The Diffuser then uses the {\color{black}\textit{inversion about the mean approach}} \cite{grover1996fast} to increase the probability amplitude of the target state through inverting  all amplitudes around their average to increase the chance of finding the correct state, which is given as
\begin{IEEEeqnarray}{lCr} 
 \textbf{Diffuser:} \quad \mathbf{D_{iff}} = 2|\phi\rangle \langle \phi| - \mathbf{I}, 
\end{IEEEeqnarray} 
\noindent where $|\phi\rangle = \frac{1}{\sqrt{D}} \sum_{x=0}^{D-1} |x\rangle$ and $\textbf{I}$  is the identity operator. It is important to note that  Grover's process is executed for multiple iterations in order to increase the probability of the correct state and ensure greater precision. Moreover, concerning the number of iterations, we employ the formula presented in \cite{9833942}, which is expressed as
\begin{IEEEeqnarray}{lcr}
\delta=\lfloor \frac{\pi}{4} \left(\sqrt{\frac{D}{V}}\right) \rfloor.
\end{IEEEeqnarray}  
 \noindent {\color{black}{We note that the Grover iteration count depends on the number of valid solutions \(V\), which is assumed to be known at the BS in this work, consistent with standard Grover-search analyses \cite{9833942,11059714}. In practical grant-free CD-NOMA scenarios, however, \(V\) may need to be estimated using techniques such as quantum counting, adaptive Grover search, or machine learning-based estimation methods \cite{11432642}, which we consider as part of future work.}}
To ensure we feed Grover's circuit with compatible input, the received signal $y_{BS}$ must be preprocessed before entering the reference register, which is designed for binary data. Hence, the received signal will be translated to the nearest integer and then converted to binary numbers. Assuming unipolar codes, where each user's code is derived from $\{0,1\}$ set,   the integer component of $y_{BS}$ is given as \cite{9833942}
\begin{IEEEeqnarray}{lcr}
y_I=\min\left(\max\left(0,\text{round}(y_{BS})\right), 2^m-1\right),
\end{IEEEeqnarray}  
\noindent where the number of discrete levels that the processed signal may adopt is determined by \(m\), which regulates the precision of the quantization process. {\color{black}{ This quantization process introduces another practical challenge in the proposed framework, since the oracle matching process relies on quantized received observations, which may make the AUD performance sensitive to noise and quantization errors.}}
 
\subsection{Physical Layer Security Analysis}
In the subsequent phase of our scenario,  BS will relay useful messages to authorized users and AN to eavesdroppers.  In the presence of eavesdroppers, the objective of beamforming and AN is to maximize the secrecy capacity for  the legitimate users.  Consequently, we employ the secrecy capacity as a performance metric, defined as the highest reliable communication rate (in bits/sec/Hz) necessary for secure transmission from a transmitter to a legitimate receiver amidst eavesdroppers, expressed as
\begin{eqnarray} \label{cap-eq}
S_k= \max \{\log_2\left(1+\gamma_k\right)-\log_2\left(1+\gamma_{e}^{\mathrm{coll}}\right),0\},
\end{eqnarray}
\noindent where $\gamma_{e}^{\mathrm{coll}}$ denotes the SINR of the colluding eavesdroppers attempting to decode the signal of user $k$.

\section{Simulation Results}

The proposed scenario is validated through simulations in this section. {\color{black} It is crucial to acknowledge that the performance evaluation is computationally intensive due to evaluating the quantum component   through classical simulation of the Grover circuit \cite{piron2024quantum}, which is sufficient for the algorithmic benchmarking and secrecy-rate analysis conducted herein. Deployment on real quantum hardware, particularly current NISQ devices, would introduce additional constraints: for an $N$-user AUD problem, the circuit requires $\mathcal{O}(N)$ qubits, and the oracle depth scales with the spreading factor $\kappa$ and the quantization precision $m$; furthermore, gate errors and decoherence would degrade oracle fidelity and reduce the probability of correctly identifying the active set. Characterizing and mitigating these hardware-induced imperfections is an important open challenge for future work.} Hence,  to ensure the practicality of our methodology while maintaining its validity, we evaluate a simplified network architecture. 
The simulation parameters are set to the following values: $N=5$, $\kappa=4$, and the codes are selected as follows: $C_1=[1,0,1,0]$, $C_2=[0,1,0,1]$, $C_3=[1,1,0,0]$, $C_4=[0,0,1,1]$, 
$C_5=[1,1,1,0]$, $P_k=30$~dB, $P_{BS}=30$~dB, $V=1$, $m=3$, $A=5$, $\alpha=0.8$, $\sigma_k^2=\sigma_e^2=1$, and $\lambda_l=0.1$. {\color{black}{Note that $V=1$ denotes the assumption of a unique valid activity pattern in the search space, not a single active user; the number of active users $K$ is determined by the number of ones in the recovered binary vector $\mathbf{a}$.}} Each simulation result represents the average over $10^5$ independent iterations. In each iteration, we compute the secrecy rate of every active legitimate user as given in~(\ref{cap-eq}). The reported "average secrecy rate" in Figs.~2--4 is obtained by averaging $S_k$ over all active users and over all Monte Carlo realizations.

\begin{figure}  
  \centering
  \includegraphics[width=0.9\linewidth]{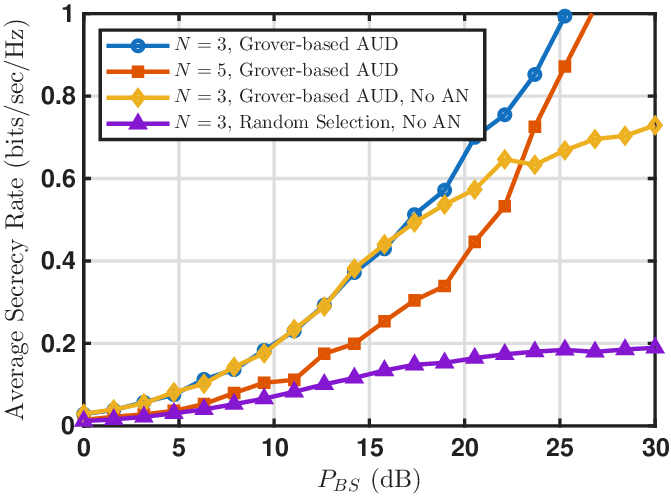}
  \caption{\color{black}{Average secrecy rate versus  $P_{BS}$ for different number of users $(N)$ and active users selection approaches. }} 
  \label{fig2}
\end{figure}

Fig. \ref{fig2} depicts the average secrecy capacity in relation to   $P_{BS}$.  When the BS transmits at a higher $P_{BS}$, the average secrecy rate increases due to enhanced legitimate link quality. Moreover,  as $N$ increases, the privacy of the shared messages declines due to two main reasons. First, Grover-based AUD requires $O(\sqrt{2^N})$ iterations; hence, larger $N$ makes detection more error-prone under noise, impacting the effective legitimate set. Second, as $N$ grows, the expected number of detected actives $(K)$ increases, which reduces the AN null-space dimension from roughly $A-K$ and thereby weakens jamming effectiveness.  We also compare the case of AUD without AN for $N=3$ to show  significant improvement of beamforming-based AN on the secrecy capacity. Grover-based AUD is also compared to a \textit{Random Selection} baseline, where it performs worse as user selection is unrelated to the true activity pattern. Since the BS randomly chooses users to serve, it often allocates power to idle users while ignoring some actual active ones. This misalignment degrades the legitimate link quality and increases signal leakage toward eavesdroppers.

    \begin{figure}  [t]
  \centering
  \includegraphics[width=0.9\linewidth]{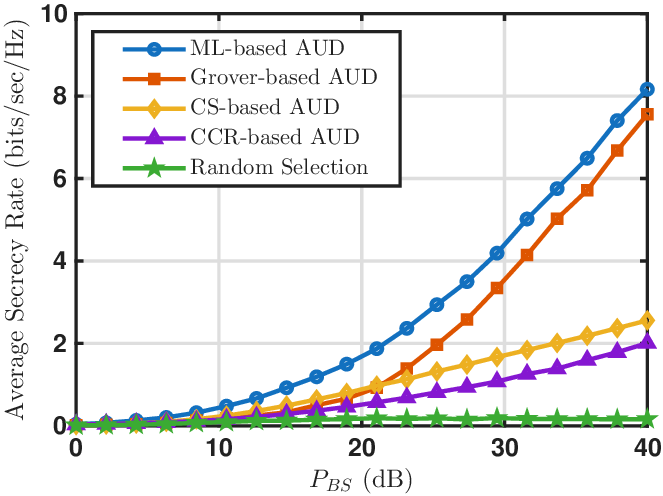}
  \caption{\color{black}{A comparison between Grover, ML, CS, CCR ($\epsilon=25$), and Random selection for AUD.     }}
  \label{fig3}
\end{figure}

Fig.~3 compares  five policies: ML AUD, {\color{black}{Grover-based AUD}}, compressive sensing (CS) AUD, classical correlation receiver (CCR) AUD, and random selection.   ML AUD is optimal  as it selects the active set $\mathbf{a}$ to minimize the Euclidean distance between the received vector $y_{\mathrm{BS}}$ and the superposition of the signatures of the selected users, expressed as
\begin{equation}
    \hat{\mathbf{a}} = \arg\min_{\mathbf{a} \in \{0,1\}^N} \left\| y_{\mathrm{BS}} - \sum_{n=1}^{N} a_n \sqrt{P_n} \, h_{\mathrm{UL},n} \mathbf{C}_n \right\|_2^2.
\end{equation}
Due to its deterministic nature and optimality, ML AUD consistently outperforms Grover-based AUD in terms of secrecy rate. However, ML still requires an exhaustive search over $2^N$ activity patterns, which becomes prohibitive as $N$ grows, whereas Grover reduces the search complexity to $\mathcal{O}(\sqrt{2^N})$ while remaining close to ML performance in the considered regime. Additionally, CCR AUD detects activity via correlation and an $\epsilon$-based thresholding criterion, making it simpler but more sensitive to noise and multiuser interference than Grover and ML approaches. CS naturally fits the AUD problem due to the sparsity of the activity vector and has also been applied in~\cite{8961111}. By jointly reconstructing the sparse activity pattern from the received signal, CS-based AUD generally outperforms CCR, which detects users individually and is therefore more vulnerable to interference and false alarms. However, CS remains sensitive to model mismatches. Finally, random selection is the least effective method, as the decision on activity is conducted independently of the true user activity.
For example, at $P_{\mathrm{BS}} = 30$~dB, {\color{black}{Grover-based AUD}} achieves an average secrecy rate within about $21\%$ of ML while outperforming CS and CCR by approximately $76\%$ and $173\%$, respectively. Here, the 21\% gap refers to
\((S_{\text{ML}} - S_{\text{Grover}})/S_{\text{ML}}\), while the improvements over CS and CCR are computed as
\((S_{\text{Grover}} - S_{i})/S_{i}\), for $i\in\{\text{CS}, \text{CCR}\}$.

\begin{figure}  
  \centering
  \includegraphics[width=1.0\linewidth]{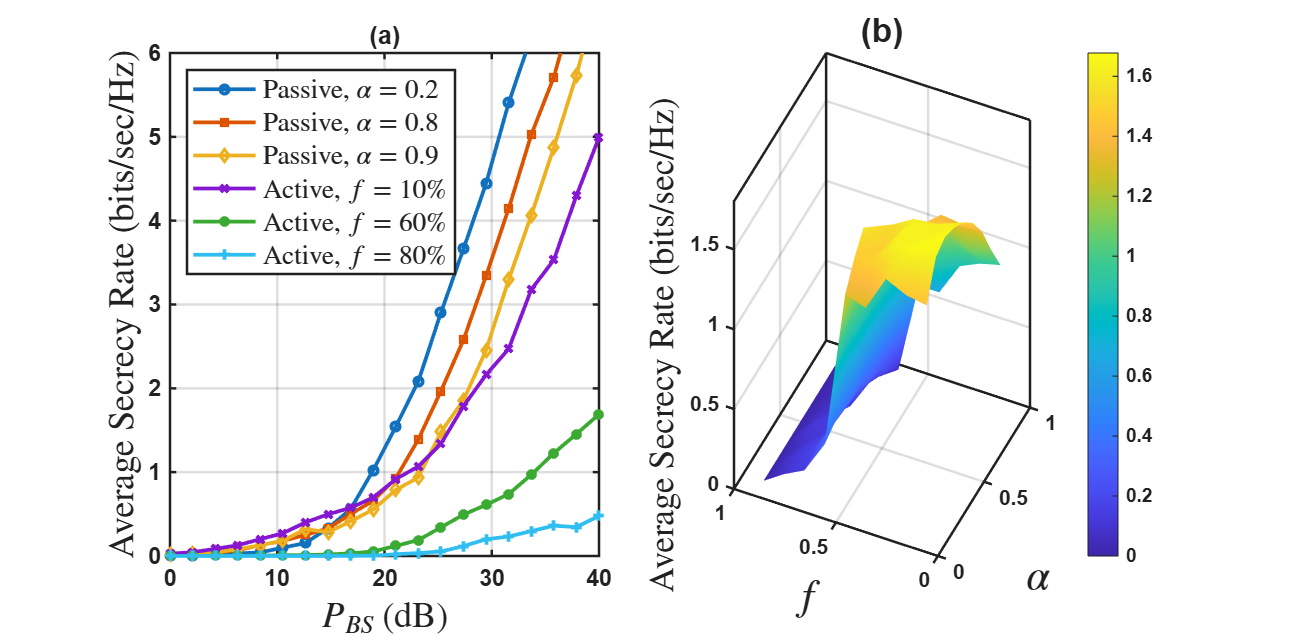}
  \caption{\color{black}{(a) Passive vs. active eavesdropper scenarios. (b) Concavity of secrecy as a function of $\alpha$ and $f$.}}
  \label{fig4}
\end{figure}

Fig.~\ref{fig4}(a) depicts the influence of the beamforming power-splitting parameter $(\alpha)$ and the selection of the top $f\%$ of the most frequent transmitters, referred to as the \textit{Active} scenario, on  performance. The impact of $\alpha$ reflects the trade-off between allocating power to information transmission and AN generation. Specifically, decreasing $\alpha$ allocates more power to AN, which strengthens jamming against eavesdroppers, whereas increasing $\alpha$ enhances the legitimate information signal while reducing the effectiveness of AN. Hence, secrecy performance depends on balancing legitimate-link enhancement and eavesdropper suppression. Moreover, increasing $f$, i.e., classifying a larger fraction of frequent transmitters as eavesdroppers, degrades secrecy as the aggregate wiretap SINR increases. The degradation is more pronounced in the \textit{Active}-eavesdropper scenario, as active users occupy part of the beamforming subspace, reducing the null space available for AN. With fewer spatial degrees of freedom, AN is less effective at suppressing eavesdroppers and may also leak interference to legitimate users. Fig.~\ref{fig4}(b) presents  3D secrecy surface versus $\alpha$ and $f$. The results show that secrecy varies concavely with $\alpha$. In particular, the highest secrecy is obtained in the region of low-to-moderate $\alpha$ and low $f$, where the trade-off between information power and jamming is best balanced.

\section{Conclusions}
This letter presents an enhancement of the PLS for a CD-NOMA network by beamforming and AN techniques through Grover-based AUD  to identify the active set of users and, subsequently, those posing a threat.  Based on Grover's detection outcome, we establish two categories of eavesdroppers: idle users and active users who transmit more frequently than others.   Our findings indicate that the average secrecy rate is improved while employing {\color{black}{Grover-based AUD}} as opposed to CS, CCR, and random selection. Compared with the optimal ML detector, Grover is slightly less accurate but offers a compelling trade-off by reducing the combinatorial search.    Furthermore, we analyze the two categories of eavesdroppers and conclude that security is less compromised when eavesdroppers remain silent.  Finally, in a three-dimensional graph, we demonstrate the feasibility of attaining a zone of elevated secrecy rate by choosing the optimal beamforming power splitting factor and the top $f\%$ most frequent transmitters.   {\color{black} Beyond AUD, the proposed Grover-based framework may also be extended to other wireless detection and estimation tasks involving large combinatorial search spaces, such as sparse signal reconstruction, beam selection, and channel estimation problems. In particular, regression- and estimation-based wireless problems requiring the identification of sparse or high-dimensional patterns may benefit from the quantum-search acceleration offered by Grover-based techniques.}


\vfill
\end{document}